# Target & Sources Accelerator Technology R&D

Snowmass2021 Topical Group Report from AF07: Targets and Sources

## Conveners:

Charlotte Barbier[a], Frederique Pellemoine[b], Yine Sun[c]

[a]Oak Ridge National Lab, Oak Ridge, TN 37831
[b]Fermi National Lab., Batavia, IL 60510
[c]Argonne National Lab., Lemont, IL 60439

1.  Targets

For the next multi-megawatt accelerator generation, targets and other beam-intercepting components (beam dumps and collimators, for instance) will face even more severe challenges due to the higher power densities, higher energy, and higher radiation. These targets will have a shorter life, more radiation damage, and harder to handle and dispose of. Designing a reliable target is already a challenge for MW class facilities today and has led to operation at limited power. Only a few facilities worldwide offer beams for target testing, and the beam provided may not be appropriate for the specific facility or project. Thus, a comprehensive research and development program must be implemented to address the challenges that multi-MW targets face. International collaborations such as RaDIATE or High Power Targetry Workshop should be leveraged and expanded to accelerate knowledge gain and avoid duplicated efforts. The next generation of high power targets will need novel designs allowing better high heat flux cooling methods, novel materials, advanced simulations, and better instrumentation. In preparation for the Summer 2022 Snowmass workshop, the target communities submitted almost twenty Letters of Intent, and four main R&D tracks were identified [1-4] and discussed below.

1.1 Irradiation Facilities

Knowing the mechanical and thermal properties of the target materials is essential to designing reliable targets. The thermo-mechanical simulations that estimate the target mechanical performance strongly depends on these parameters. However, as the beam interacts with the matter, these properties are affected due to radiation damage and cannot be predicted other than by measuring them experimentally. To measure the impact of radiation damage, low-energy ion beams are often used to irradiate samples in nuclear physics. The HEP community is looking at those beams for emulating high-energy proton irradiation [1]. The disadvantages of such approach are the poisoning of the samples with ion beam, the inability to create gas through transmutation, and the shallow penetration; in comparison with high-energy proton irradiation which penetrates deeper and creates gas through transmutation. Few high-energy proton irradiation facilities exist, such as BLIP at BNL or HiRadMat at CERN. Still, they will not be sufficient to cover the required irradiations and examinations for the next generation of high-power targets. Thus, the focus of the next 10 years should be on the development and the validation of alternative beams to simulate high-energy proton irradiation. Currently, an R&D cycle including irradiated material takes up to 4 to 5 years for single studies when they are done in collaboration with outside facilities that have the dedicated hot cell for material science. Facilities (hot cells) should be developed in accelerator facilities to examine and characterize irradiated materials on-site and shorten the R&D cycle.

## 1.2 Modeling Needs

Simulations play a critical role in Targetry component design (i.e. target as well as beam dumps, collimator, absorbers, and scrapers). They are used to predict the beam-matter interactions and the thermo-mechanical response induced by beam heating. Better interfaces between different codes must be developed to allow design exploration that will optimize new high-heat flux cooling methods and the target particles' output. Although the current particle code can predict radiation damage with variables such as DPA (Displacement per Atom) in the molecular lattice, DPA cannot be experimentally measured. There is a strong need to develop numerical tools that emulate radiation damage from the atomic to macroscopic scale [2]. Such effort has been initiated in the RaDIATE collaboration, but further work is needed to predict the physio-mechanical properties of the irradiated materials. A successful code could significantly reduce R&D cycle duration and allow the exploration of novel materials for targets.

Several groups have reported that Machine Learning (ML) has been a powerful tool for diagnostic systems: it can find hidden correlations among a vast variance parameter phase and predict a specific physics parameter. The ML effort should be continued for future multi-MW beam facilities

## 1.3 Novel materials and concepts

Conventional materials already have limitations in sustaining today's accelerator environment. There is a pressing need to explore novel materials that are more radiation damage and thermal shock resistant, while fulfilling the physics requirements [3]. Possible candidates are High-Entropy Alloys, Electrospun nanofiber materials, SiC-coated graphite and SiC-SiC composites, toughened Fine-grained Recrystallized (TFGR) tungsten, dual-phase titanium alloys, and advanced graphitic materials. Modeling and experiments are both required to validate these materials before using them in operation. Novel targetry concepts and technologies will need to be explored as the current high-power target technologies, such as liquid metal or rotating targets will face limitations and challenging costs. Advanced fabrication such as 3D-printed and advanced cladding will need to be developed to get better cooling. Other target concepts such as flowing granular should also be explored.

## 1.4 Radiation hardened instrumentation

Today's targets are generally poorly instrumented due to high radiation or safety concerns (like introducing new material in the beam). As the accelerator gets more powerful, target designs will get more complex with a shorter lifespan. There will be a strong desire to keep the target operating as long as possible to keep the accelerator operating cost reasonable. Since simulations cannot accurately predict the target's life, there will be a strong need for radiation-hardened instrumentations to monitor the target's health [4]. These sensors would also be used to validate in-depth the numerical tool used to emulate the target and potentially leveraged advanced machine learning methods.

## 1.5 Conclusion

All the four topics above are critical to fully support the development of the next generation (near and longer-term) accelerator. However, the priority should be on reducing the R&D cycle handling radioactive materials by developing in-house capabilities. More funding and efforts should be spared for the targetry R&D as it becomes more and more challenging.

The RaDIATE collaboration, a US high-energy proton irradiation initiative, needs to be expanded and more supported financially by GARD (DOE HEP General Accelerator R&D) to not only improve our understanding of radiation damage but also to explore novel materials and bring the target modeling to the needed level for high power accelerator. Synergy with R&D on Accelerator Driven Systems and with nuclear Energy and Nuclear Physics communities should be leveraged since they are facing similar challenges.

The key recommendations for the Target Accelerator Technology R&D are:

- Develop or upgrade Post-Irradiation Examination (PIE) facilities that will allow testing activated novel materials.
- Develop and validate alternative beams to simulate high-energy proton irradiation.
- Develop irradiation stations relevant to assess radiation damage in various material.
- Develop and validate numerical tools to predict radiation damage.
- Develop novel concept and novel target material.
- Develop radiation-resistant sensors for better numerical validation of the structural and thermal simulations.
- Strengthen the high-power target community through the RaDIATE collaboration and High-Power Targetry workshop.

2. Electron sources [5]

    2.1. Cathodes
        2.1.1. Highly polarized electron sources are required as probes for high energy physics. Developing a steady, robust, reliable source of the hetero-structure based cathode such as GaAs etc. in the near-term is of utmost importance. Photocathode material beyond III-V semiconductors should be explored [6]. Increasing cathode lifetime for higher average currents is critical and will require exploration of cathode technologies, developments in electron guns and drive laser technology.
        2.1.2. Unpolarized electron source with high quantum efficiency, small intrinsic emittance is another area for cathode development. Research towards future cathode technologies calls for a collaborative theoretical and experimental effort between the fields of materials science, condense matter physics and accelerator physics.
        2.1.3. Development of cathode testing facilities is critical in testing the performance of novel photocathode candidate materials in electron guns under realistic operating conditions.

    2.2. Guns
        DC, NCRF and SCRF guns all have particular strengths. DC guns have best vacuum for GaAs photocathodes and generate highest CW current, while NCRF guns reach higher gradient essential for low emittance and high brightness, and SCRF guns are promising with vacuum levels comparable to DC guns but at higher gradients.
        2.2.1. DC guns: higher operating voltage is desired to mitigate space-charge forces. Maintaining vacuum conditions to preserve the GaAs-based photocathode life time, and managing ion back-bombardment at high CW current are active and highly relevant areas for R&D.
        2.2.2. NCRF guns: higher accelerating gradient is desired. Copper structures operated in cryogenic temperatures show potential for both higher gradient and repetition rate. Higher average current is more manageable in lower frequency RF guns.
        2.2.3. SCRF guns: normal conducting photocathode insertion into a SCRF gun is a challenging aspect in the SCRF gun technology. Collaborations with leading labs (such as HZDR and BNL) should be encouraged for the development of SCRF guns in USA, such as the LCLS-II-HE injector SCRF gun injector project.

2.3. Injectors

The electron injector delivers beam up to 100 MeV. The injectors can be either CW or pulsed.

2.3.1. Achieving high 6D phase space brightness, mitigating deleterious collective effects and developing methods to manipulate the phase space distribution of the electron beam to precisely fit applications are the core priorities of injector R&D. This includes photo-injector drive-laser shaping, mitigation of space charge and coherent synchrotron radiation, round-to-flat beam transformation, emittance exchange between transverse and longitudinal phase space, pulse train generation, etc.

2.3.2. Reducing dark current is one of the most important topics for CW electron injectors.

2.4. Advanced concepts

LWFA and PWFA based electron sources focused on the generation of unpolarized electrons are promising sources and have demonstrated impressive beam parameters in the past decade but still need substantial development to serve as mature drivers for HEP colliders.

3. Ion sources [7]

3.1. Electron Cyclotron Resonance Ion Source

The critical challenge in advancing the state of ECRIS lies in achieving higher magnetic field to support operation at higher heating frequency and to provide stronger plasma confinement. The future development of ECRIS should go forward:

3.1.1. Development of a 4th generation ECRIS. The R&D work is to develop a magnet system that can produce the needed magnetic minimum-B field configuration with either NbTi or Nb3Sn conductor. This is the most urgent task that needs sufficient funding to move forward right now.

3.1.2. If the MARS magnet geometry gets validated in ECRIS, an HTS or a hybrid (HTS + other conductors) based MARS magnet capable of generating magnetic fields for operating at ~ 80 GHz will be the pathway to a future generation super-performing ECRIS.

3.1.3. After the issue of higher field magnets has been overcome, higher frequency microwave generation with suitable power level in cw mode for ECRIS operations could be another important task and that should be addressed.

3.2. Laser Ion Source

The laser ion source is a simple ion source that can provide intense pulsed beams. Here is the list for future R&D:

3.2.1. High repetition rate operation. The laser power density must be increased to obtain high charge states, but these results in cratering of the target, and the target supply cannot keep up during high repetition rate operation. To solve this problem, tape targets can be prepared and wound at high speed, or liquid targets can be prepared.

3.2.2. Long pulse widths. The pulse width depends on the velocity distribution of ions in the generated plasma. If the velocity at which the center of gravity of the plasma moves can be reduced, the time difference within the pulse of the plasma reaching the extraction field can be increased. To achieve this, a foil with a thickness of less than 1 µm could be used as a target. However, in this case, the plasma would not have enough initial velocity and would be expected to diffuse in all directions. To prevent this, ionization in a high magnetic field and limiting the direction of diffusion may prevent the plasma density from decreasing.
3.2.3. Generation of high charge-state ions from heavy elements. Optimize the laser irradiation conditions for the species and charge state. The laser pulse width survey would be a useful strategy to have optimum laser irradiation conditions.
3.2.4. Instability of the ion beam is often caused by the roughness of the target surface. To prevent this, the beam can be further stabilized by adjusting the lens position while precisely measuring the position of the target and focus lens.

3.3. Charge breeders

Future RIB facilities are expected to significantly increase the production rates of rare isotopes by several orders of magnitude. Such facilities may necessitate the implementation of the two systems, EBIST and ECRIS breeders, for complementarity in charge breeding RIB of different intensity ranges (ECRIS: high-intensity, EBIST: low-intensity). Effort to improve their performance both in parallel is therefore important.

3.3.1. R&D in high current, high current-density electron guns for EBIST breeders is needed for efficient charge breeding of high-intensity beams. This would enable two operating modes: 1) pulsed injection at high repetition rates to remain within the charge-capacity limit of the upstream cooler-buncher and 2) continuous injection with longer breeding times for beam rates exceeding the buncher's capacity. Such guns would also allow for the reach of lower A/Q values with higher efficiencies with shorter breeding times, extending the energy limit of post-accelerated beams to isotopes having shorter half-lives. A long-term target goal of 4 A and ~5500 A/cm2 is suggested.

3.3.2. The pervasive background of ECRIS breeders can substantially contaminate RIB of low intensity. The development of new techniques and designs to reduce it should be considered a priority.

3.3.3. ECRIS breeders generally operate with a single and fixed microwave heating frequency leading to a fixed element-dependent charge-state distribution. This limits the energy range of the post-accelerated beams. Fine-frequency tuning, two-frequency heating, and the afterglow effect can allow for low charge states (short breeding times) to be optimized to maximize the breeding efficiency of short-lived isotopes, extending with these isotopes the energy limit of post-accelerated beams from ECRIS breeders. R&D is ongoing and needed in this direction.

3.3.4. Constant progress in numerical simulation tools and advances in computing power is enabling the development of increasingly detailed self-consistent numerical simulations codes for ECRIS and EBIST breeders. The development of complete EBIST-and ECRIS-breeder 3D numerical simulation codes (full 3D geometries) is highly necessary to improve their performance (increase efficiencies and provide beams of higher quality) and should be set as a (continuous) long-term goal. This appears accessible in the future with increasingly faster computers and the availability of computing centers.

3.4. Machine learning and Ion sources

The optimization of an ion source's operation space, some tens of parameters, is difficult and time-consuming for a human. The application of machine learning to this task promises advancement in both source performance and the maintenance of stability.

3.4.1. Optimization of ion sources via machine learning is still a young area of research. It should be expected that efforts directed at the application of machine learning to ion source operation would result in real successes in the not-too-distant future.

4. Positron Source

High intensity polarized position sources are critical to future e$^+$e$^-$ colliders. A coherent effort in the US accelerator physics community is needed. Opportunities for advanced technology in positron sources are reported in a white paper in AF06: Advanced accelerator concepts [8].

5. Acknowledgements

The authors would like to thank organizers and participants in the Snowmass2021 Electron Source Workshop [9], members of the Snowass2021 Ion Source Working Group [10] participants of the target related workshops [11,12,13], authors of the white papers and LOIs for their hard work and contribution to the content presented in this working group summary.

[1] Pellemoine, F., et al. "Irradiation Facilities and Irradiation Methods for High Power Target." *arXiv preprint arXiv:2203.08239* (2022).

[2] Barbier, C., et al. "Modeling Needs for High Power Target." *arXiv preprint arXiv:2203.04714* (2022).

[3] Ammigan, K., et al. "Novel Materials and Concepts for Next-Generation High Power Target Applications." *arXiv preprint arXiv:2203.08357* (2022).

[4] Yonehara, K. "Radiation hardened beam instrumentations for multi-Mega-Watt beam facilities." *arXiv preprint arXiv:2203.06024* (2022).

[5] Daniele Filippetto, Joe Grames, Carlos Hernandez-Garcia, Siddharth Karkare, Philippe Piot, John Power, Yine Sun and Erdong Wang, "Electron Sources for Accelerators," Snowmass2021, Accelerator Frontier, AF7, https://arxiv.org/abs/2207.08875.

[6] Luca Cultrera. "Spin polarized electron beams production beyond III-V semiconductors", https://arxiv.org/abs/2206.15345.

[7] Alain Lapierre, Janilee Benitez, Masahiro Okamura, Damon Todd, Daniel Xie, and Yine Sun, "Ion Sources for Production of Highly Charged Ion Beams," Snowmass2021, Accelerator Frontier, AF7. https://arxiv.org/abs/2205.12873.

[8] P. Musumeci, C. Boffo, S. S. Bulanov, I. Chaikovska, A. Faus Golfe, S. Gessner, J. Grames, R. Hessami, Y. Ivanyushenkov, A. Lankford, G. Loisch, G. Moortgat-Pick, S. Nagaitsev, S. Riemann, P. Sievers, C. Tenholt, K. Yokoya, "Positron Sources for Future High Energy Physics Colliders," https://arxiv.org/pdf/2204.13245.

[9] Snowmass2021 Electron Source Workshop, Organized by Yine Sun (Chair), Joe Grames, Siddharth Karkare, Daniele Filippetto, Carlos Hernandez-Garcia, John Power, and Erdong Wang, Feb. 16-18, 2022, https://indico.fnal.gov/event/46053/.

[10] Snowmass2021 Ion Source Working Group: Yine Sun (convener), Alain Lapierre (Technical Lead), Edward Beebe, Janilee Benitez, Masahiro Okamura, Damon Todd and Daniel Xie.

[11] Snowmass Target Simulations Workshop, Organized by Charlotte Barbier, April 6 2021, https://conference.sns.gov/event/267/

[12] Snowmass2021 Irradiation Stations and Alternatives, Organized by Frederique Pellemoine, June 17-18 2021, https://indico.fnal.gov/event/48628/

[13] Snowmass2021 Device health monitoring (integrated system + ML) and Rad-Hard instrumentation workshop, Organized by Katsuya Yonehara, February 3-4 2022, https://indico.fnal.gov/event/52543/